\documentstyle[12pt]{article}

\begin{document}

\begin{flushright}
IMSc/2000/06/24 \\ 
cond-mat/0006279
\end{flushright} 

\vspace{2mm}

\vspace{2ex}

\begin{center}
{\large \bf Tsallis Statistics: Averages and a Physical} \\ 

\vspace{2ex}

{\large \bf  Interpretation of the Lagrange Multiplier $\beta$} \\

\vspace{8ex}

{\large  S. Kalyana Rama}

\vspace{3ex}

Institute of Mathematical Sciences, C. I. T. Campus, 

Taramani, CHENNAI 600 113, India. 

\vspace{1ex}

email: krama@imsc.ernet.in \\ 

\end{center}

\vspace{6ex}

\centerline{ABSTRACT}
\begin{quote} 
Tsallis has proposed a generalisation of the standard entropy,
which has since been applied to a variety of physical systems.
In the canonical ensemble approach that is mostly used, average
energy is given by an unnromalised, or normalised,
$q$-expectation value. A Lagrange multiplier $\beta$ enforces
the energy constraint whose physical interpretation, however, 
is lacking. Here, we use a microcanonical ensemble approach 
and find that consistency requires that only normalised
$q$-expectation values are to be used. We then present a
physical interpretation of $\beta$, relating it to a physical
temperature. We derive this interpretation by a different method
also.
\end{quote}

\vspace{2ex}


\newpage

\vspace{4ex}

{\bf 1.}  
Tsallis has proposed \cite{tsallis1} a one parameter
generalisation of the standard Boltzmann-Gibbs entropy.  
It is given by 
\begin{equation}\label{sq} 
S_q = \frac{\sum p_i^q - 1}{1 - q} 
\end{equation} 
where $p_i$ is the probability that the system is in a state
labelled by $i$, the sum, here and in the following, runs over
all the allowed states, and the parameter $q$ is a real number.
We have set the Boltzmann constant $k$ equal to unity.  In the
limit $q \to 1$, $S_q = - \sum p_i ln p_i$, thus reducing to the
standard Boltzmann-Gibbs entropy. The statistical mechanics that
follows from the entropy $S_q$, referred to here as Tsallis
statistics, is rich in applications and has been studied
extensively. It retains the standard thermodynamical structure,
leads to power-law distributions as opposed to the exponential
ones that follow from the standard statistical mechanics, and
has been applied to a variety of physical systems. See
\cite{tsallis,spl} for a thorough discussion, and \cite{brazil}
for an exhaustive list of references.

Mostly, in these applications, a canonical ensemble approach is
used \cite{tsallis1,123}. The entropy $S_q$ is extremised
subject to the constraint $\sum p_i = 1$, and an average energy
constraint (equation (\ref{123}) below), obtaining thus the
probability distribution $p_i$. Various thermodynamical
quantities are then calculated by standard methods\footnote{
Since the standard thermodynamical structure is preserved, such
quantities can be calculated using the standard formulae
\cite{tsallis}. However, they are mathematical constructs only,
which may or may not have a physical meaning.}.

The averages can be taken to be given by unnormalised, or 
normalised, $q$-expectation values. Calculations are simpler
with the former choice, whereas the later choice has all
desireable properties. Also, various quantities calculated using
these two choices, although not equal to each other, can be
related by a set of formulae. Therefore, both choices have
often been used. See \cite{123} and references therein for a
detailed discussion. The average energy constraint is then
enforced through a Lagrange multiplier, $\beta$. In the standard
statistical mechanics, $\beta$ is the inverse of the temperature
which can be physically measured. To the best of our knowledge,
a similar physical interpretation of $\beta$ in Tsallis
statistics is still lacking. (However, see the Note at the end
of the paper.)

In the present letter, we use the microcanonical ensemble
approach \cite{tsallis1} and calculate the temperature and
specific heat using the entropy $S_q$.  For classical ideal gas,
various thermodynamical quantities have been calculated in
\cite{prato,abe} using the canonical ensemble approach. Upon
comparing the specific heats obtained in the microcanonical and
the canonical ensemble approaches, we find that these two
approaches can be equivalent, and various quantities calculated
in these two approaches can be equal to each other, only if the
average is given by normalised $q$-expectation value, and not by
unnormalised one.

We then present a physical interpretation of the temperature
calculated in the microcanonical ensemble. Comparing then with
the canonical ensemble results of \cite{abe}, we obtain a
physical interpretation of the Lagrange multiplier $\beta$.
Interestingly, the same interpretation can be derived by a
simple refinement of the Gibbsian argument of Plastino and
Plastino \cite{plastino}. We present this derivation also.

The plan of the paper is as follows. We first give the relevant
details \cite{123} and results \cite{prato,abe} of the canonical
ensemble approach. We then present our results, and close with a
few comments.

{\bf 2.} 
In the canonical ensemble approach, the entropy $S_q$ is
extremised subject to the constraint $\sum p_i = 1$, and an
average energy constraint. One thus obtains the probability
distribution $p_i$, and calculates the physical quantities by
standard methods. We briefly present here the relevant
aspects. We follow \cite{123}, where more details can be found.

The average energy $U$ of the system can be 
taken to be given by unnormalised, or 
normalised, $q$-expectation value as follows: 
\begin{equation}\label{123} 
U_2 = \sum p_i^q \epsilon_i 
\; , \; \; \; {\rm or} \; \; \; 
U_3 = \frac{\sum p_i^q \epsilon_i}{\sum p_i^q} \; , 
\end{equation}
where $\epsilon_i$ is the $i^{th}$ state energy. Here and in the
following, the subscript $2$ ($3$) indicates that the averages
are given by unnormalised (normalised) $q$-expectation values.
Extremising $S_q$, with the average energy given by $U_2$
($U_3$) in (\ref{123}), gives
\begin{eqnarray}
p_{i 2} & = & \frac{a_{i 2}}{Z_2} 
\; , \; \; \; 
a_{i 2} = \left( 1 - \beta_2 (1 - q) \epsilon_i
\right)^{\frac{1}{1 - q}} \label{pi2} \\
p_{i 3} & = & \frac{a_{i 3}}{Z_3} 
\; , \; \; \; 
a_{i 3} =  \left( 1 - \frac{\beta_3 (1 - q) 
(\epsilon_i - U_3)}{Y_3} \right)^{\frac{1}{1 - q}} 
\; , \label{pi3} 
\end{eqnarray} 
where $Z_{2(3)} = \sum a_{i 2(3)}$, and 
$\beta_2$ ($\beta_3$) 
is the Lagrange multiplier for $U_2$ ($U_3$).
It also follows that 
\[
Y_2 \equiv \sum p_{i 2}^q = 
Z_2^{1 - q} + \beta_2 (1 - q) U_2 \; , \; \; \; 
Y_3 \equiv \sum p_{i 3}^q = Y_3 = Z_3^{1 - q} \; . 
\] 

{\bf 3.} 
Consider systems whose standard Boltzmann-Gibbs,
namely $q = 1$, partition function is of the form
\begin{equation}\label{zbg}
Z_{BG} \propto l^a \beta^{- a} 
\end{equation}
where $a$ is a dimensionless parameter, $l$ is a characteristic
length, and $\beta$ is the inverse temperature. For example, for
a $d$-dimensional classical ideal gas with $N$ particles and
volume $V$, $a = \frac{d N}{2}$ and $l \propto V^{\frac{1}{d}}$.
For classical ideal gas, the above formalism has been applied,
and various quantities such as energy, specific heat, etc. have
been calculated in \cite{prato,abe}. The average energies 
$U_2, U_3$ and the specific heats ${\cal C}_2, {\cal C}_3$ are
given by \footnote{ As mentioned in section {\bf 1}, 
$(\beta_2, U_2, {\cal C}_2, \cdots)$ and 
$(\beta_3, U_3, {\cal C}_3, \cdots)$ can be related to each
other by a set of formulae \cite{2to3}.}
\begin{eqnarray}
\beta_2 U_2 = \frac{a Y_2}{1 + (1 - q) a} 
\; , & \; \; \; & {\cal C}_2 = a Y_2 \label{betac2} \\
\beta_3 U_3 = a Y_3 \; , & \; \; \; & 
{\cal C}_3 =  \frac{a Y_3}{1 - (1 - q) a} \; . \label{betac3} 
\end{eqnarray} 
Note that the specific heats ${\cal C}_2$ and ${\cal C}_3$ are
not equal to each other, and even have qualitatively different
behaviour: ${\cal C}_2$ is always positive since $Y_2 \equiv
\sum p_{i 2}^q$ is always positive, whereas ${\cal C}_3$ becomes
negative when $(1 - q) a > 1$.  Explicit expressions for $Y_2$
and $Y_3$ can be found in \cite{prato,abe}, but are not needed
here.

{\bf 4.}  
However, one can also use the microcanonical ensemble approach
\cite{tsallis1}. For a given energy $E_{mc}$ of the system, 
the entropy is extremised subject only to the constraint 
$\sum p_{i (mc)} = 1$. The subscript $mc$, 
here and in the following, 
indicates that the microcanonical ensemble approach is 
used. Note that no averaging of the energy is involved in 
this approach and, hence, no choice is made.

The entropy $S_q$ is extremised in the case of equiprobability,
{\em i.e.} when all the probabilities are equal, with the
extremum being a maximum (minimum) if $q > 0$ ($q < 0$)
\cite{tsallis1}. Thus, if $W$ is the number of allowed 
states then 
\begin{equation}\label{sqmc}
p_{i (mc)} = \frac{1}{W} 
\; \; \; {\rm and,} \; \; {\rm hence,} \; \; \; 
S_q = \frac{W^{1 - q} - 1}{1 - q} \; . 
\end{equation}
Since the standard thermodynamical structure is preserved 
\cite{tsallis}, 
the inverse temperature $\beta_{mc}$ and the specific heat 
${\cal C}_{mc}$ can be calculated using the standard 
formulae: 
\begin{eqnarray}
\beta_{mc} & \equiv & \frac{\partial S_q}{\partial E_{mc}} 
= \beta_* Y_{mc} \label{beta*} \\
{\cal C}_{mc} & \equiv & - \beta_{mc}^2 \left( 
\frac{\partial^2 S_q}{\partial E_{mc}^2} \right)^{- 1} 
= \frac{c_* Y_{mc}}{1 - (1 - q) c_*} \; , \label{c*}
\end{eqnarray}
where we have defined 
$Y_{mc} = \sum p_{i (mc)}^q = W^{1 - q}$, 
\begin{equation}\label{std}
\beta_* = \frac{\partial ln W}{\partial E_{mc}} 
\; , \; \; \; {\rm and} \; \; \; 
c_* = - \beta_*^2 \left( 
\frac{\partial^2 ln W}{\partial E_{mc}^2} \right)^{- 1} \; . 
\end{equation}
For systems whose $q = 1$ partition function is given by
(\ref{zbg}), we have 
\begin{equation}\label{sbg}
W \propto l^a E_{mc}^a \
\; \; \; {\rm and,} \; \; {\rm hence,} \; \; \; 
\beta_* = \frac{a}{E_{mc}} 
\; , \; \; \; {\rm and} \; \; \; 
c_* = a \; . 
\end{equation}
Equations (\ref{beta*}) and (\ref{c*}) then give 
\begin{equation}\label{betac}
\beta_{mc} E_{mc} = a Y_{mc} \; , \; \; \; 
{\cal C}_{mc} = \frac{a Y_{mc}}{1 - (1 - q) a} \; . 
\end{equation} 

{\bf 5.}
For systems whose $q = 1$ partition function is given by
(\ref{zbg}), we now have expressions (\ref{betac2}),
(\ref{betac3}), and (\ref{betac}) for energy and specific 
heat, obtained using the canonical \cite{prato,abe} and the
microcanonical ensemble approach. The canonical ensemble
approach involves an averaging of the energy, and the
expressions (\ref{betac2}) ((\ref{betac3})) are for the case
where the average is given by unnormalised (normalised)
$q$-expectation value. However, in the microcanonical ensemble
approach, no averaging of energy is involved and, hence, no
choice is made.

Let us now compare the specific heats. The specific heats given
by (\ref{betac3}) and (\ref{betac}) are identical, upto factors
involving $Y_3$ and $Y_{mc}$, and differ distinctly from that
given by (\ref{betac2}). For example, the specific heat given by
(\ref{betac2}) is always positive, whereas the specific heats
given by (\ref{betac3}) and (\ref{betac}) become negative when
$(1 - q) a > 1$. Therefore, it follows that the microcanonical
and the canonical ensemble approaches can be equivalent, and
various quantities calculated in these two approaches can be
equal to each other, only if the average is given by normalised
$q$-expectation value, and not by unnormalised one. 

{\bf 6.}
From now on, we assume that the averages are given by
normalised $q$-expectation values only. Therefore, energy and
specific heat in the canonical ensemble approach are given by
(\ref{betac3}). We now present a physical interpretation of
$\beta_*$ in (\ref{std}) and, thus, also of $\beta_{mc}$ 
and $\beta_3$. 

Consider a system obeying Tsallis statistics, with $q$ positive
but otherwise arbitrary, with energy $E$ and number of allowed
states $W(E)$, and enclosed within a container with which it can
exchange energy only. Together, let them be isolated. 
Thus, if $E_c$ is the energy of the container then the
total energy $E_{tot} = E + E_c$ is fixed. Let the container be
choosen to obey the standard Boltzmann-Gibbs statistics,
namely Tsallis statistics with $q = 1$. Therefore, if
$W_c(E_c)$ is the number of allowed states of the container,
then 
\begin{equation}\label{betaphys}
\beta_{phys} = \frac{\partial ln W_c}{\partial E_c} 
\end{equation} 
is the inverse of its temperature, which can be physically 
measured. 

We would like to find the values $E$ of the system, and 
$E_c = E_{tot} - E$ of the container, at which the (system +
container) is in equilibrium. But the analysis of such a
composite system, where the constituent systems have different
values of $q$, is highly nontrivial and is still an open
problem\footnote{We thank the referee for emphasising this
point.}. Nevertheless, it is reasonable to expect that 
{\bf (i)} the entropy of the composite system is
extremised in the case of equiprobability;
{\bf (ii)} the extremum is a maximum, at least in the
case where the $q$'s of the constituent systems are all
positive; and
{\bf (iii)} the maximum is a monotonically increasing
function of the total number of allowed states of the composite
system. 

Although we are unable to justify these properties rigorously,
they appear to be physically reasonable, and are satisfied by
any single system obeying Tsallis statistics with $q > 0$
\cite{tsallis1}, see equation (\ref{sqmc}). Hence, we assume
that {\em any composite system, the $q$'s of whose constituent
systems are all positive, also satisfies the properties {\bf
(i)-(iii)} given above.}  Since the entropy is extremised in
equilibrium, our assumption then implies that {\bf (i')} when a
composite system is in equilibrium, the energies of its
constituents will be such as to maximise the total number of
states of the composite system\footnote{ Alternatively, we may
instead assume that {\em the composite system satisfies the
property {\bf (i')} only}, which will suffice for our purposes
here.}. Note that no assumption is made, or implied, about the
explicit form of the entropy of the composite system by assuming
the properties {\bf (i)-(iii)} or {\bf (i')} for the composite
system.

In the case of the (system + container) considered above, this
assumption then implies that in equilibrium, the energy $E$ of
the system, and the energy $E_c = E_{tot} - E$ of the container,
will be such as to maximise the total number of states of the
(system + container), given by 
\[
W_{tot}(E_{tot}) = W(E) W_c(E_{tot} - E) \; . 
\]
Hence, with $E_{tot} = E + E_c$ fixed, we have that 
in equilibrium, 
\begin{equation}\label{pathria}
\frac{\partial ln W(E)}{\partial E} = 
\frac{\partial ln W_c(E_c)}{\partial E_c} \; . 
\end{equation} 
It then follows from equations (\ref{beta*}), 
(\ref{std}), and (\ref{betaphys}) that 
\begin{equation}\label{big0}
\beta_{phys} = \beta_* = \frac{\beta_{mc}}{Y_{mc}} \; , 
\end{equation}
which relates $\beta_{mc}$ of the microcanonical
ensemble approach to $\beta_{phys}$, the physical inverse
temperature of the container. As clear from its derivation, 
the above relation is valid for any arbitrary system,  
whose $q = 1$ partition function is completely general. 

Now, assuming the equivalence of the microcanonical and the
canonical ensemble approach, we can set $E_{mc}(\beta_{mc}) =
U_3(\beta_3)$. For systems considered here, whose $q = 1$
partition function is given by (\ref{zbg}), it then follows from
equations (\ref{betac3}), (\ref{betac}), and (\ref{big0}) that 
\begin{equation}\label{big}
\frac{\beta_3}{Y_3} = \frac{\beta_{mc}}{Y_{mc}} = 
\beta_{phys}  \; , 
\end{equation} 
which relates $\beta_3$ to $\beta_{phys}$, the inverse
temperature of the container, which can be physically measured.
Equation (\ref{big}) thus provides a physical interpretation of
the Lagrange multiplier $\beta_3$ of the canonical ensemble
approach.

{\bf 7.}
The relation (\ref{big}) between $\beta_{phys}$ and $\beta_3$
can also be derived by another method. Plastino and Plastino
have derived the probability distribution of the form 
given in (\ref{pi2}), for $q < 1$, by a Gibbsian argument
\cite{plastino}. A simple refinement of their argument 
leads to the probability distribution of the form given 
in (\ref{pi3}). Requiring it to be exactly identical 
with (\ref{pi3}) then leads to (\ref{big}).

The argument of \cite{plastino} is, briefly, the following.
Consider a large, but finite, heat bath obeying the standard
Boltzmann-Gibbs thermodynamics. Let $E_b$ be its energy and
$\beta_{phys}$ its inverse temperature, which can be physically
measured. Also, let the total number of states in the energy
range $(E_b \pm \frac{\Delta}{2})$ be $\eta (E_b) \Delta$. 
Consider now a system weakly interacting with such a heat
bath. Then, the probability $p_i (\epsilon_i)$ that the system
is in a state $i$, with energy $\epsilon_i$, is given by 
\[
p_i (\epsilon_i) \propto \eta (E_b - \epsilon_i) \Delta \; . 
\]
Assuming that 
$\eta (E) \propto E^{\alpha - 1}$, where $\alpha \gg 1$,
one obtains
\begin{equation}\label{pipp}
p_i (\epsilon_i) \propto \left( 1 - \frac{\epsilon_i}{E_b} 
\right)^{\alpha - 1} \; . 
\end{equation}
Also, $\beta_{phys} = \frac{\alpha - 1}{E_b}$. 
Let $q = \frac{\alpha - 2}{\alpha - 1}$. Then, $q < 1$, 
$\alpha - 1 = \frac{1}{1 - q}$, and 
\[
p_i (\epsilon_i) \propto \left( 1 - \beta_{phys} (1 - q) 
\epsilon_i \right)^{\frac{1}{1 - q}} \; , 
\]
which is of the form given in (\ref{pi2}). 

By a simple refinement of the above argument, one can obtain the
probability distribution of the form given in (\ref{pi3}). The
above expression for $\beta_{phys}$ assumes that the heat
bath always has energy $E_b$, irrespective of the energy of 
the system. However, the actual energy of the heat
bath is $E_b - \epsilon_i$ when the system is in state $i$. 
Therefore, if the average energy of the system is $U$ then
the average energy of the heat bath is $E_b - U$. Hence, 
a more precise expression for $\beta_{phys}$ is given by  
\[
\beta_{phys} = \frac{\alpha - 1}{E_b - U} \; .
\]
Using this expression in (\ref{pipp}), and with $q$ 
defined as above, one obtains 
\begin{equation}\label{pi3pp}
p_i (\epsilon_i) \propto \left( 1 - \beta_{phys} (1 - q) 
(\epsilon_i - U) \right)^{\frac{1}{1 - q}} \; , 
\end{equation}
which is of the form given in (\ref{pi3}). Requiring this 
distribution to be identical with (\ref{pi3}) then gives 
\[
\beta_{phys} = \frac{\beta_3}{Y_3} \; , 
\]
which is the same relation as in (\ref{big}) and relates
$\beta_3$ of the canonical ensemble approach to $\beta_{phys}$,
the physical inverse temperature of the bath. Assuming 
the validity of the argument of \cite{plastino} and our 
refinement of it, the above relation is valid for any 
arbitrary system, but with $q < 1$. 

{\bf 8.} 
We close with a few comments. The normalised $q$-expectation
values have, indeed, been found earlier to possess desireable
properties and, hence, considered to be the appropriate
ones. Here, we find that this result follows simply by
requiring the equivalence between the microcanonical and the
canonical ensemble approaches.

However, we have considered here only systems whose $q = 1$
partition function is given by (\ref{zbg}). Hence, it is
desireable to establish this result for any arbitrary system
whose $q = 1$ partition function is completely general.

The relation (\ref{big0})
between $\beta_{mc}$ and $\beta_{phys}$ is valid
for any arbitrary system whose $q = 1$ partition function is
completely general. The relation (\ref{big}) 
between $\beta_3$ and
$\beta_{phys}$ is derived, in the first method, only for
systems whose $q = 1$ partition function is given by
(\ref{zbg}). Assuming the validity of the argument of
\cite{plastino} and our refinement of it, the second method 
of derivation is valid for any arbitrary system, but with 
$q < 1$. Hence, a general relation between $\beta_3$ and
$\beta_{phys}$, valid for any arbitrary system and for any 
value of $q$, is still lacking.

Also, Tsallis statistics is applied to a variety of diverse
physical systems such as Levy flights, turbulence, etc. to name
but a few \cite{tsallis,spl}. It is not clear if each one
of them can be modelled as a system within a container, or as 
a system weakly interacting with a large, but finite, heat 
bath - models which played a crucial role in the physical
interpretation of $\beta_3$ presented here. On the other hand,
however, one may instead assume that $\frac{\beta_3}{Y_3}$,  
or a suitable generalisation of it, is
indeed a physical quantity as given in (\ref{big}). 
Its study may then, perhaps, provide
new insights into physical systems, to which Tsallis
statistics is applied.

\vspace{3ex}

{\bf Note:} 
While this work was being written, a paper 
by Abe et al \cite{recent} has appeared. In the 
prescription termed {\em optimal Lagrange multipliers} 
formalism \cite{olm} which they use, the combination 
$\frac{\beta_3}{Y_3}$ appears naturally. As shown in 
\cite{recent}, certain key properties of the ideal gas 
then become identical in both the standard statistical 
mechanics and Tsallis statistics. 

The referee has brought to our attention a paper by Abe
\cite{aberef} where also the combination $\frac{\beta_3}{Y_3}$,
termed {\em a renormalised (inverse) temperature},
appears naturally while establishing the zeroth law of
thermodynamics using the classical ideal gas model. 

\vspace{3ex}

{\bf Acknowledgement:} 
We are grateful to G. Baskaran for introducing us to Tsallis
statistics. We thank G. Baskaran, G. I. Menon, P. Ray, and
B. Sathiapalan for many discussions. Also, we thank the referees
for their comments, suggestions for improvement, and 
for bringing \cite{aberef} to our
attention.

%\newpage


\begin{thebibliography}{999}
\bibitem{2to3}
The probability distributions in (\ref{pi2}) and (\ref{pi3}) are
related by $\beta_2 = \beta_3 \left( Y_3 + \beta_3 (1- q) U_3
\right)^{- 1}$. Then $p_{i 2} = p_{i 3}$ and, therefore, 
$Y_3 = Y_2$, $U_3 = \frac{U_2}{Y_2}$, and $\beta_3 =
\frac{\beta_2 Y_2^2}{Z_2^{1 - q}}$. Other quantities such as
free energy, specific heat, etc.  can also be correspondingly
related \cite{123}.
\end{thebibliography}
\end{document}